\documentstyle[aps,multicol,prl,epsfig,floats]{revtex}

\begin{document}

\twocolumn[\hsize\textwidth\columnwidth\hsize\csname@twocolumnfalse%
\endcsname

\title{Disorder Driven Critical Behavior of Periodic Elastic Media in a
Crystal Potential}

\author{Jae Dong Noh and Heiko Rieger}

\address{Theoretische Physik, Universit\"{a}t des Saarlandes, 66041
Saarbr\"{u}cken, Germany}

\maketitle

\begin{abstract}
  We study a lattice model of a three-dimensional periodic elastic
  medium at zero temperature with exact combinatorial optimization
  methods. A competition between pinning of the elastic medium,
  representing magnetic flux lines in the mixed phase of a
  superconductor or charge density waves in a crystal, by randomly
  distributed impurities and a periodic lattice potential gives rise
  to a continuous phase transition from a flat phase to a rough
  phase. We determine the critical exponents of this roughening
  transition via finite size scaling obtaining $\nu\approx1.3$,
  $\beta\approx0.05$, $\gamma/\nu\approx2.9$ and find that they are
  universal with respect to the periodicity of the lattice potential.
  The small order parameter exponent is reminiscent of the random
  field Ising critical behavior in 3$d$.
\end{abstract}


]
  
A number of materials possess an instability towards the formation of
a periodically modulated structure in space below a particular
temperature. Prominent examples are charge-density wave systems
\cite{cdw}, where a Peierls instability leads to a state with
periodically varying charge modulation, or magnetic flux-lines in the
mixed phase of high-temperature superconductors \cite{blatter}, where
the long-range interaction among the lines results in the formation of
the Abrikosov flux-line lattice. Other systems forming such periodic
structures are spin density waves in anisotropic metals, polarization
density waves in incommensurate ferroelectrics, and mass density waves
in superionic conductors \cite{waves}.

Usually the periodicity ${\bf q}$ of this state of broken
translational invariance, is incommensurate with the underlying
crystal lattice, but if ${\bf q}$ and a reciprocal lattice vector
${\bf k}$ become commensurate (${\bf q}\approx{\bf k}/p$) the density
wave locks in at this wave vector.  If fluctuations --- either thermal
or induced by impurities, i.e.,\ quenched disorder --- are weak, these
systems are then in a {\it flat} phase that can be quantified by an
order parameter reflecting the broken symmetry. Deviations of the
local density from the perfect periodic structure can be measured by a
displacement field $\phi({\bf r})$, which shows long range order in
the flat phase. When the fluctuations become too strong this
long-range order vanishes at a roughening transition and the system
enters a {\it rough} phase in which the displacement-displacement
correlations $\langle[\phi({\bf r})-\phi({\bf 0})]^2\rangle$ diverge with
the distance $r$.

In the presence of quenched disorder this roughening transition is
driven by the the competition between two pinning forces acting on the
periodically modulated flat phase: one coming from the underlying
lattice potential preferring long-range order, the other by the point
impurities tending to destroy it. The universality class of this
transition in the experimentally most relevant case of three space
dimensions (3$d$), its critical exponents and the scaling laws, have
not been directly scrutinized up to now and are the topic of this
paper.  Thermal fluctuations are expected to be (dangerously)
irrelevant at the roughening transition \cite{emig}, and the critical
behavior at the transition should be dominated by a zero temperature
fixed point analogous to random field critical behavior
\cite{villain}.  Hence the universal properties of the
roughening transition at {\it finite} temperatures are expected to be
identical to the one at {\it zero temperature} and the critical
exponents can in principle be extracted numerically by exact ground
state calculations \cite{ko-review,rfim-ground}, which is the method
that we will use here.

The model Hamiltonian that captures the universal properties of the
roughening transition under consideration should contain the following
features: It should be formulated in terms of a (scalar) displacement
field $\phi({\bf r})\in(-\infty,+\infty)$; an elastic energy term
$\frac{\gamma}{2} (\nabla \phi)^2$ as the first order (elastic)
approximation of the interaction energy arising from small
deformations of the flat state $\phi({\bf r})={\rm const.}$; a periodic
potential $V_{\rm per}(\phi)=V_{\rm per}(\phi+2\pi/p)$ modeling the
crystal lattice; and a random potential $V_{\rm rand}(\phi)$ mimicking
the effect of impurities, which should be invariant under the global
shift of the whole displacement field $\phi\to\phi+2\pi$. 
The commensurability parameter $p$ entering the
periodic potential is integer for the lock-in state and is given by
the ratio of lattice constant of the elastic media with respect to
that of the underlying periodic potential. The following Hamiltonian
fulfills these requirements \cite{waves,emig}:
\begin{equation}\label{H_con}
{\cal H} = \int d^d {\bf r} \left[ \frac{\gamma}{2} | \nabla \phi |^2
- v \cos (p \phi) + \eta  \cos( \phi - \varphi )\right] \,
\label{hamil}
\end{equation}
where $\varphi({\bf r})$ are independent quenched random variables
uniformly distributed on $[-\pi,\pi]$ and $\gamma$, $v$, and
$\eta({\bf r})$ denote the elastic constant, the periodic potential
strength, and the random potential strength, respectively.  The
underlying elastic approximation for this model is valid as long as
disorder induced topological defects do not proliferate. In 2$d$ this
actually happens \cite{defects2d}, but in 3$d$ the elastic medium is
stable for weak disorder \cite{defects3d}.

For $v=0$, the Gaussian variational and the functional renormalization
group~(FRG) calculations~\cite{korshunov,Giamarchi_Doussal94} and numerical
studies \cite{mcnamara} show that the system is in the elastic glass
phase, corresponding to a zero-temperature fixed point, at all
temperatures.  The elastic glass phase in 3$d$ is
characterized by diverging fluctuations
\begin{equation}\label{G_lnr}
G({\bf r}) = \overline{ \langle [\phi({\bf r_0}+{\bf r})) - 
             \phi({\bf r_0})]^2 \rangle } \simeq 
             2 A \ln | {\bf r} |
\end{equation}
at large distances with a universal coefficient $A$. The overbar
denotes the disorder average and $\langle\ldots\rangle$ the spatial
average over ${\bf r}_0$ and the thermal average.

A simple scaling argument \cite{larkin,imry-ma,emig} shows that for
$d>2$ the flat phase ($\phi={2\pi n}/{p}$, with $n$ a fixed integer)
is stable as long as the disorder is weak enough: For vanishing
disorder $\eta=0$ an excitation $\phi\to\phi+2\pi/p$ over a terrace of
linear scale $\xi$ costs an elastic energy of the order of
$\xi^{d-1}$, whereas for non-vanishing disorder the same excitation
could gain energy of order $\xi^{d/2}$.  Thus for $d>2$ the elastic
energy loss will dominate over weak disorder and the ground state
stays flat. Only a strong enough disorder will drive the periodic
medium into the rough phase. This disorder driven roughening
transition was first \cite{aharony} studied within a variational theory 
in \cite{bouchaud}, where a first order transition was found,
whereas the FRG method used in \cite{emig} predicted
a continuous roughening transition for $p >
p_c(d)=6/(\pi\sqrt{\epsilon})$ with $\epsilon=4-d$ at finite disorder
strength that is determined by a zero-temperature fixed
point. The order parameter exponent $\beta$ and the correlation length
exponent $\nu$ were given to leading order in a double expansion in
$\epsilon$ and $\mu={p^2}/{p_c^2}-1$ by
\begin{equation}\label{RG_result}
\nu^{-1} = 4\mu 
\quad, \quad 
{\beta}/{\nu} = ({\pi^2}/{18})\,\epsilon \ .
\end{equation}
A naive insertion of $d=3$ and $p_c=6/\pi$ into these expressions
yields values for $\beta$ and $\nu$ that are incompatible with our
results which we report now.

We consider a discrete model for the continuum Hamiltonian~(\ref{hamil}).
Due to the periodic potential the elastic medium remains flat on a 
microscopic length scale with
\begin{equation}\label{phi_h}
\phi = ({2\pi}/{p})\,h \quad , \quad h \mbox{ integer} \ .
\end{equation}
Therefore, on a coarse-grained level, the medium can be described by
this integer height variable $\{h_{\bf x}\}$ representing a
$(3+1)$-dimensional surface on a simple cubic lattice with sites ${\bf
x}\in \{1,\ldots,L\}^3$. Creating steps costs elastic energy and the
surface is subjected to a random pinning potential. These effects plus
the periodic potential are incorporated in the following
solid-on-solid~(SOS) model Hamiltonian
\begin{eqnarray}
{\cal H} &=& \sum_{({\bf x,y})} J_{(h_{\bf x},{\bf x}); (h_{\bf y},{\bf y})}
                | h_{\bf x} - h_{\bf y} |  \nonumber \\
         &-& \sum_{{\bf x}} \eta_{\bf x} \cos [ (2\pi / p) h_{\bf x} -
           \varphi_{\bf x} ] \ , 
\label{H_discrete}
\end{eqnarray}
where the first sum runs over nearest neighbor pairs $({\bf x},{\bf
y})$ on a simple cubic lattice. The Hamiltonian has to be be invariant
under a global shift $h \rightarrow h+p$, which is inherited from the
symmetry under $\phi\rightarrow \phi+2\pi$ of the continuum
Hamiltonian (\ref{hamil}). Hence we impose a periodicity in the step
energies $J$ by $J_{(h+p,{\bf x});(h'+p,{\bf y})} = J_{(h,{\bf
x});(h',{\bf y})}$.  Although various aspects of the microscopic
physics of the lattice model (\ref{H_discrete}) and the continuum
Hamiltonian (\ref{hamil}) might be different, we can expect the
roughening transition occurring in both models to be in the same
universality class, since both models have identical
symmetries. This 3$d$ SOS model is then mapped onto a (3+1)$d$
ferromagnetic random bond Ising model with an anti-periodic boundary
condition in the extra-direction, denoted by $u$ \cite{2dper}. In the
ground state the latter induces a 3$d$ interface, identical to the
surface we are searching (up to a global shift) that can be determined
{\it exactly} by using a min-cut/max-flow algorithm (see
\cite{ko-review,mcnamara} for details).

\begin{figure}[t]
  \centerline{\epsfig{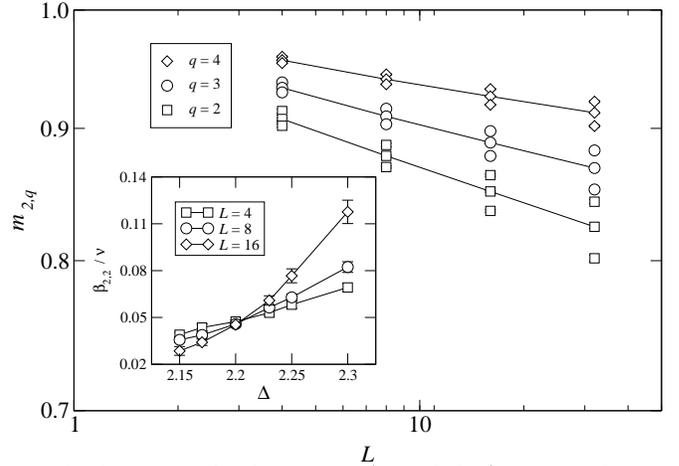}}
\caption{Magnetizations $m_{p,q}$~($q=2,3,4$) at $p=2$ and $\Delta=2.17,
  2.20,$ and $2.23$ from top to bottom. The data at the critical point
  $\Delta_c = 2.20$ are connected by solid lines. The inset shows the
  effective exponents for
  $m_{2,2}$~(Eq.~(\ref{beta_eff})).}\label{magP2}
\end{figure}

We study the model (\ref{H_discrete}) with $p=2,3,$ and $4$ in finite
$L^3\times U$ lattices with $L\le 32$. The lattice size in the $u$
direction, $U$, has to be chosen large enough in order to avoid an
interference of the surface with the boundary.  Random couplings are
assigned to the bonds in a unit cell of size $L^3\times p$ and they
are repeated periodically in the $u$ direction. Then the exact ground
state configuration is calculated using a max-flow algorithm.  We present
numerical results obtained by using the uniform distribution for
$0\leq \varphi_{\bf x} < 2\pi$ and $ 0 \leq \eta_{\bf x} < V$ and the
exponential distribution, $P(J) = J_0^{-1}e^{-J/J_0}$, for $J>0$.  The
results we report do not depend on the choice of the distribution.
The strength of the random pinning potential is denoted by $\Delta
\equiv {V}/{J_0}$ and we will vary this quantity to trigger the
roughening transition in our system.

For each $p$, we measure the magnetizations 
\begin{equation}
m_{p,q}(L,\Delta) = \overline{ | \langle 
e^{2\pi i h_{\bf x}/q} \rangle|} \quad (q=2,3,\ldots) 
\end{equation}
with $\langle (\cdots) \rangle$ and $\overline{(\cdots)}$ denoting the
spatial and the disorder average, respectively, in the ground state.
Typically the disorder average is taken over $10000 \sim 3000$ samples
for $L=4 \sim 32$.  Note that the order parameter $m=\overline{\langle
  e^{i\phi}\rangle}$ considered in Ref.~\cite{emig} corresponds to
$m_{p,q=p}$, cf.,~Eq.~(\ref{phi_h}).  

In Fig.~\ref{magP2} we show the magnetization for
$p=2$ as a function of $L$, which scales at the
critical point $\Delta=\Delta_c$ like $m_{p,q} \sim
L^{-\beta_{p,q}/\nu}$, with $\beta_{p,q}$ and $\nu$ the order
parameter and correlation length exponent, respectively. This scaling
is followed best by the data at $\Delta=2.20$, whereas there
is a downward~(upward) curvature for $\Delta = 2.23~(2.17)$ when
plotting $\ln m_{p,q}$ vs. $\ln L$. The critical point $\Delta_c$ can
be directly determined by looking at the effective exponent
\begin{equation}
[\beta_{p,q} / \nu ]_{L} \equiv - 
          \frac{\ln(m_{p,q}(2 L) / m_{p,q}(L))} {\ln 2} \ ,
\label{beta_eff}
\end{equation}
which is (asymptotically) independent of system size at the critical point
and equal to the critical exponents $\beta_{p,q}/\nu$~(see Table~I).

\begin{figure}[t]
\centerline{\epsfig{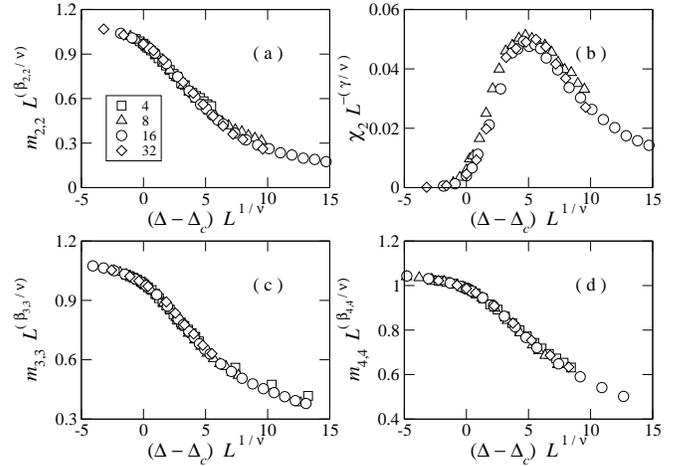}}
\caption{Scaling plots of the magnetization for $p=2$ (a), $p=3$ (c), 
  $p=4$ (d), and the susceptibility for $p=2$ with
  $\gamma/\nu=2.90$ (b).  Parameters used in the plots are
  listed in Table~I.}
\label{scaling_plot}
\end{figure}

The correlation length exponent is obtained from the scaling behavior
near the critical point.  Each quantity is a function of 
$L / \xi$ with the correlation length $\xi \sim |\Delta-\Delta_c|^{-\nu}$
such that the scaling form of the magnetization is
\begin{equation}\label{m_scale}
m_{p,q}(L,\Delta) = L^{-\beta_{p,q} / \nu} \ 
                    {\cal F}((\Delta-\Delta_c) L^{1/\nu}) 
\end{equation}
with a scaling function ${\cal F}$.  Using the values of $\Delta_c$
and $\beta_{2,2}/\nu$ estimated previously, we determine the
correlation length exponent as the optimal value which yields the best
data collapse of $m_{2,q=2}(L,\Delta)$.  The estimated correlation
length exponent is also listed in Table~I and the scaling
plot is given in Fig.~\ref{scaling_plot} (a).

\begin{table}
\caption{Estimates for the critical exponents for different
commensurability parameter $p$ obtained via finite size scaling from 
the numerical data.}\label{table1}
\begin{tabular}{c|cccccc}
   & $\Delta_c$ & $\beta_{p,2}/\nu$ & $\beta_{p,3}/\nu$ & $\beta_{p,4}/\nu$  & 
          $\nu$       \\
\hline
$p=2$ &  2.20(3)   & 0.046~(5) & 0.034~(3) & 0.022~(3) & 1.25(5)   \\
$p=3$ &  2.475(25) & 0.049~(7) & 0.037~(9) & 0.024~(4) & 1.29(5) \\
$p=4$ &  2.95(5)   & 0.044~(5) & 0.033~(5) & 0.022~(5) & 1.28(8) \\
\end{tabular}
\end{table}

The correlation length exponent is also determined from 
the susceptibility defined as
\begin{equation}
\chi_p  = 
L^3 \left( \overline{ | \langle e^{2\pi i h_{\bf x}/p} \rangle |^2 } -
          \overline{ | \langle e^{2\pi i h_{\bf x}/p} \rangle | }^2 \right)
\ .
\end{equation}
Near the transition point it develops a peak, whose position scales as
$(\Delta^*(L) - \Delta_c ) \sim L^{1/\nu}$ and whose height as
$\chi^*(L) \sim L^{\gamma/\nu}$ with the susceptibility exponent
$\gamma$. For each $L$, $\Delta^*$ and $\chi^*$ are obtained by
fitting the susceptibility curve near the peak with a quadratic
function, and then the critical exponents are extracted to yield that
$\nu^{-1} = 0.76(5)$ and $\gamma/\nu = 2.90(5)$ for $p=2$.  
Both estimates of
$\nu$ from the magnetization and the susceptibility are consistent
with each other, and the susceptibility exponent satisfies the scaling
relation, $\gamma/\nu = d - 2 \beta_{2,2}/\nu$ within the error bars.
Fig.~\ref{scaling_plot} (b) shows the scaling plot of $\chi_2
L^{\gamma/\nu}$ versus $(\Delta - \Delta_c) L^{1/\nu}$ with $\Delta_c
= 2.20$, $\nu = 1.25$, and $\gamma/\nu=2.90$. Except for the smallest
system size $L=4$, those exponents collapse the data well.

We have performed the same analysis for $p=3$ and $p=4$ and present
the critical points and the critical exponents in Table~I.
Fig.~\ref{scaling_plot} (c) and (d) show the corresponding scaling
plots of $m_{p=3,q=3}$ and $m_{p=4,q=4}$.

The order parameter exponents are so small that they might also be
interpreted as a signature of a first order transition with $\beta=0$.
However, they deviate from zero systematically for all $p$ and $q$
and, furthermore, the scaling analyses show that the transition is
associated with the diverging correlation length. Therefore we
conclude that the transitions are of second order with small but
finite order parameter exponents, which is reminiscent of the critical
behavior of the 3$d$ random-field Ising model~(RFIM)~\cite{rfim-ground,RFIM},
for which $\beta/\nu = 0.012(4)$~\cite{rfim-ground}.

Our numerical results deviate from the FRG results~\cite{emig} in many
respects. The critical exponents are substantially different from the
analytic results of $\nu\simeq 2.59$ and $\beta_{2,2}/\nu \simeq
0.548$ for $p=2$~(see Eq.~(\ref{RG_result})). Moreover, the order parameter
and the correlation length exponents appear to be independent
of $p$ within the error bars.

\begin{figure}[t]
\centerline{\epsfig{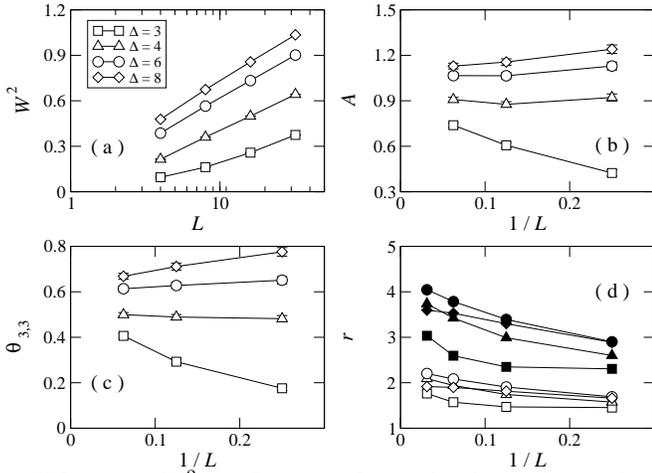}}
\caption{(a) $W^2$ vs. $L$ in semi-log scale, (b) prefactor $A$ of $W^2 \sim
  A \ln L$, (c) effective exponent $\theta_{3,3}$ for $m_{3,3} \sim
  L^{-\theta_{3,3}}$, and (d) exponent ratios $r =
  \theta_{3,2}/\theta_{3,3}$~(empty symbols) and
  $\theta_{3,2}/\theta_{3,4}$~(filled symbols). All data are made from
  $p=3$ in the disordered phase.}
\label{scaling_rough}
\end{figure}

The discrepancy between FRG and the numerical results is surprising
since the FRG method produces the correct scaling of the elastic glass
phase, i.e., the logarithmic growth of fluctuations as in
Eq.~(\ref{G_lnr}). Then the roughness $W^2 \equiv (2\pi /p)^2 \left[
  \overline{ \langle h_{\bf x}^2\rangle - \langle h_{\bf x}\rangle^2}
\right] = \frac{1}{2L^3} \sum_{{\bf x}} G({\bf x})$ also grows
logarithmically in $L$ as $W^2 \simeq A \ln L$ with
$A=\frac{\pi^2}{9}(4-d)$ \cite{Giamarchi_Doussal94}.  These results
are confirmed numerically for a SOS model in 3$d$ \cite{mcnamara},
where $W^2 \simeq A \ln L$ is found with a prefactor $A\simeq 1.0$ in
accordance with the analytic prediction. We also find this agreement
for our model (\ref{H_discrete}) in the glass
phase ($\Delta>\Delta_c$):
Fig.~\ref{scaling_rough} (a) shows the logarithmic scaling of the
roughness $W^2$.  The prefactor $A$ is estimated as $[W^2(2L) -
W^{2}(L)] / \ln{2}$ and plotted in Fig.~\ref{scaling_rough} (b). The
estimates have a strong $L$-dependence.  Due to lack of a theory for
corrections to scaling, we fit them with a quadratic polynomial in
$1/L$ to extrapolate the asymptotic value of $A$. The results are
$0.98 \lesssim A \lesssim 1.11$, consistent with the analytical and
the numerical results. If the fluctuations of the
elastic glass phase have the Gaussian nature as assumed in
Ref.~\cite{Giamarchi_Doussal94}, the roughness, $W^2\sim A\ln L$, and
the magnetizations, $m_{p,q} \sim L^{-\theta_{p,q}}$, are not
independent quantities since they should obey $\overline{\langle
  e^{i\phi({\bf r})-i\phi({\bf r'})}\rangle} \sim e^{-\frac{1}{2}
  \overline{\langle [\phi({\bf r}) - \phi({\bf r'})]^2\rangle}}$
implying the relation
\begin{equation}\label{Gscaling}
A = 2~ \theta_{p,p} \quad , \quad \theta_{p,q}/\theta_{p,q'} = q'^2 / q^2  
\ .
\end{equation}
Fig.~\ref{scaling_rough} (c) shows the effective exponents
$\theta_{3,3}$. The polynomial fitting is used to extrapolate the asymptotic
value, which yields $0.51 \lesssim \theta_{3,3} \lesssim 0.62$. We also
calculate ${\ln m_{3,2}}/{\ln m_{3,3}}$ and 
${\ln m_{3,2}}/{\ln m_{3,4}}$, which approach 2.25 and 
4.0, respectively, as $L$ increases, cf.,~Fig.~\ref{scaling_rough} (d).
Those values satisfy the scaling relations in Eq.~(\ref{Gscaling}) 
approximately. This is a strong evidence for the Gaussian nature of the
fluctuations in the elastic glass phase and hence justifies the analytic
approaches in this regime.

However, when using the same procedure as in the glass phase to
determine $A_c$ at the critical point for $p=2,3,4$, we obtain $A_c
\simeq 0.18~(p=2), 0.092~(p=3)$, and $0.046~(p=4)$. Note that $A_c$
appears to be approximately inversely proportional to $p^2$, which
implies that the bare width $W_0^2 \equiv \overline{ \langle{h_{\bf
      x}^2}\rangle - \langle{h_{\bf x}}\rangle^2}$ is independent of
$p$. The Gaussian theory requires that
\begin{equation}
A_c = 2 ( \beta_{p,p}/\nu ) \quad ,\quad \beta_{p,q}/\beta_{p,q'} = q'^2/q^2
\label{gausspred}
\end{equation}
but $A_c$ and $(\beta_{p,p}/\nu )$ listed in Table~I violate the first
relation by a factor of two. The ratios $\beta_{p,2}/\beta_{p,3}\simeq
1.5$ and $\beta_{p,2}/\beta_{p,4}\simeq 2.3$ are also far from the
values $9/4$ and $4$, required by (\ref{gausspred}). This implies a
strongly non-Gaussian nature of the fixed point in 3$d$ and provides a
hint why the FRG-prediction (\ref{RG_result}) for 3$d$ differs from
ours. On the other hand, the latter are based on a double expansion
around $d=4$ and $p=p_c$ and it is well possible that $d=3$ and the
values for $p$ we have considered here are simply beyond the validity
of such first order perturbation expansion. We want to stress
that we think that the system sizes we studied are sufficiently large
to see the true asymptotic behavior of the roughening transition
since we capture correctly the features of the rough phase fixed
point.

In summary we presented the first numerical study of a disorder driven
roughening transition in a periodic elastic medium. Our results for
the critical exponents deviate significantly from the predictions of a
recent analytical FRG calculation and we discussed the validity of its
underlying Gaussian approximation. We found that this new universality
class is reminiscent of random field critical behavior in 3$d$
including a very small order parameter exponent.
To complete the picture of the underlying zero temperature fixed point
scenario, one has to compute the violation of hyperscaling exponent
$\theta$, which necessitates techniques different from those used in
this work~\cite{noh}.


\begin{references}

\bibitem{cdw}
   \vskip-1.3cm
   G.\ Gr\"uner, Rev. Mod. Phys. {\bf 60}, 1129 (1988).

\bibitem{blatter}
  G.\ Blatter {\it et al.}, 
  Rev. Mod. Phys. {\bf 66}, 1125 (1994).

\bibitem{waves}
  G.\ Gr\"uner, Rev. Mod. Phys. {\bf 66}, 1 (1994);
  G. Gr\"uner, {\it Density waves in solids} (Addsion-Wesley, Reading, 1994).

\bibitem{emig} 
  T. Emig and T. Nattermann, Phys. Rev. Lett. {\bf 79}, 5090 (1997); 
  Eur. Phys. J. B {\bf 8}, 525 (1999).

\bibitem{villain}
  J. Villain, J. Physique {\bf 46}, 1843 (1985).

\bibitem{ko-review} M.~J. Alava, P. Duxbury, C. Moukarzel, and H.
  Rieger, in {\em Phase Transitions and Critical Phenomena} Vol. {\bf
    18}, (ed.\ C. Domb and J. L. Lebowitz), p.141--317, (Academic
  Press, Cambridge, 2000).

\bibitem{rfim-ground}
  See e.g., A. A. Middleton and D. S. Fisher, preprint.

\bibitem{defects2d}
  C.~Zeng {\em et al.}, Phys. Rev. Lett. {\bf 82}, 1935 (1999);
  F.\ Pfeiffer and H. Rieger,
  J.Phys. A {\bf 33}, 2489 (2000);
  A. A. Middleton, 
  Phys. Rev. B {\bf 61}, 14787 (2000).

\bibitem{defects3d}
 J. Kierfeld, T. Nattermann, and T. Hwa, Phys. Rev. B {\bf 55}, 626 (1997); 
 M.~J.~P. Gingras and D.~A. Huse, Phys. Rev. B {\bf 53}, 15193 (1996);
 D.~S. Fisher, Phys. Rev. Lett. {\bf 78}, 1964 (1997).

\bibitem{korshunov}
 S. E. Korshunov, Phys. Rev. B {\bf 48}, 3969 (1993).

\bibitem{Giamarchi_Doussal94} 
 T. Giamarchi and P. Le Doussal, Phys. Rev. Lett. {\bf 72}, 1530 (1994).

\bibitem{mcnamara} 
 D. McNamara {\em et al.}, Phys. Rev. B {\bf 60}, 10062 (1999).

\bibitem{larkin}
 A. Larkin, Sov. Phys. JETP {\bf 31}, 784 (1970).

\bibitem{imry-ma} 
 Y. Imry and S.~K. Ma, Phys. Rev. Lett. {\bf 35}, 1399 (1975).


\bibitem{aharony} 
  A related magnetic Hamiltonian with random fields was studied in:
  A. Aharony and E. Pytte, Phys. Rev. B {\bf 27}, 5872 (1983).

\bibitem{bouchaud} 
  J.-P. Bouchaud and A. Georges, Phys. Rev. Lett. {\bf 68}, 3908 (1992).

\bibitem{2dper}
  A similar model in 2$d$ does {\it not} 
  have a roughening transition: M. Alava, E. Sepp\"al\"a, and P. Duxbury,
  Phys. Rev. E {\bf 62}, 3230 (2000), {\it ibid.} {\bf 63}, 036126 (2001).

\bibitem{RFIM} 
 A.~T. Ogielski, Phys. Rev. Lett. {\bf 57}, 1251 (1986);
 H. Rieger and A.~P. Young, J. Phys. A {\bf 26}, 5279 (1993);
 H. Rieger, Phys. Rev. B {\bf 52}, 6659 (1995); 
 M.~R. Swift {\em et al.}, Europhys. Lett. {\bf 38}, 273 (1997); 
 A.~K. Hartmann and U. Nowak, Eur. Phys. J. B {\bf 7}, 105 (1999). 

\bibitem{noh} 
 J.~D. Noh and H. Rieger, in preparation.

\end{references}
\end{document}